# DAMPING OF COMPRESSIONAL MHD WAVES IN QUIESCENT PROMINENCES AND PROMINENCE-CORONA TRANSITION REGION (PCTR)

K.A.P SINGH

*Department of Applied Physics, Institute of Technology*
*Banaras Hindu University, Varanasi-221005, India*
*and*
*Indian Institute of Astrophysics, Bangalore 560034, India*
*(email: alkendra1978@yahoo.co.in)*

## Abstract

The effects of radiative losses due to Newtonian cooling and MHD turbulence have been considered to examine the spatial damping of linear compressional waves in quiescent prominences and prominence – corona transition region (PCTR). The radiative losses give acceptable damping lengths for the slow mode wave for the radiative relaxation times in the range ($10$-$10^3$s). From prominence seismology, the values of opacity and turbulent kinematic viscosity have been inferred. It has been found that for a given value of radiative relaxation time, the high frequency slow mode waves are highly damped. We have also investigated the possible role of MHD turbulence in damping of MHD waves and found a turbulent viscosity can re-produce the observed damping time and damping length in prominences, especially in PCTR.

## 1. INRODUCTION

Prominences are masses of relatively cool ($T \sim 10^4$K) and dense ($\rho \sim 10^{-10}$ kgm$^{-3}$) material suspended in the corona ($\rho \sim 10^{-14}$ kg m$^{-3}$, $T \sim 10^6$K). The spectra of prominences hold the key to understanding their physical conditions e.g., temperatures, densities, pressures etc. The internal structure and physical properties of prominences, however, can be studied through the new tool of prominence seismology. Magnetohydrodynamic (MHD) waves and oscillations of the solar prominence have been carried out both from the ground and from space (Patsourakos and Vial, 2002). Small amplitude waves (or oscillations) with velocity amplitudes from 0.1 km s$^{-1}$ to 2-3 km s$^{-1}$ have been observed (e.g., Bashkirtsev and Mashnich 1984, Molowny-Horas *et al*., 1999). Molowny-Horas et al. (1997) observed the oscillations in different parts of the quiescent prominence and assuming the plane wave propagation, they obtained the group speed, phase speed and the perturbation wavelength. Using the VTT telescope at Sac Peak, Molowny-Horas *et al*. (1999) found velocity perturbations with periods between 28 and 95 min at different locations in a prominence and observed that the amplitude of the oscillations decreases in time with damping times between 101 and 377 min. Terradas *et al*. (2002) investigated the temporal and spatial variations of oscillations and reported strong damping of oscillations with damping times between two to three times the wave period. Terradas *et al*. (2001) considered energy losses through Newtonian cooling and found that only the slow mode waves are affected by damping leaving fast mode waves almost undamped. Carbonell *et al*. (2004) studied the damping of MHD waves in an unbounded medium by considering the effect of prominence-corona transition region (PCTR) with energy losses due to thermal conduction, optically thin radiation and heating. Terradas *et al*. (2005) showed the result also holds in a bounded medium. Recently, Ballai (2003) has reviewed some of the possible mechanisms that operate in prominences that can explain the spatial damping of linear compressional waves.

In this paper, we consider the radiative losses due to Newtonian cooling to study the spatial damping of MHD waves over a wide range of radiative relaxation times. To model the spatial damping of the prominence oscillations we consider a homogeneous equilibrium configuration which is unbounded in all the directions with magnetic field in the x direction and neglected the effect of the gravity. The radiative losses due to Newtonian cooling are invoked in an adiabatic energy equation with complex $\hat{\gamma}$ (Bunte and Bogdan, 1994), which is given by $\hat{\gamma} = \frac{1 + \mathrm{i}\omega\tau_{\mathrm{R}}\gamma}{1 + \mathrm{i}\omega\tau_{\mathrm{R}}}$. We discuss the role of MHD turbulence in the main body of the prominence and prominence-corona transition region (PCTR) and compare our results with the available observations. The energy losses by Newtonian cooling using local dispersion relation method are presented in section 2, and the MHD turbulence models are presented in section 3 while the results and discussion are given in the last section.

## 2. Energy Losses by Newtonian Cooling

Considering a homogeneous equilibrium configuration which is unbounded in all directions with magnetic field in the x-direction and neglecting the effect of gravity, we have

$$p_0 = \text{constant},\ \rho_0 = \text{constant},\ T_0 = \text{constant} \quad \boldsymbol{B}_0 = B_0\hat{\boldsymbol{x}}, \quad \mathbf{v}_0 = 0 \tag{1}$$

where $p_0$, $\rho_0$, $T_0$ and $B_0$ are equilibrium values of pressure, density, temperature and magnetic field.

The relevant MHD equations are given by:

$$\frac{D\rho}{Dt} + \rho\nabla\cdot\boldsymbol{V} = 0 \tag{2}$$

$$\rho\frac{Dv}{Dt} = -\nabla p + \frac{1}{\mu}(\nabla\times\boldsymbol{B})\times\boldsymbol{B} + \rho\,\boldsymbol{g} \tag{3}$$

$$\frac{\rho^{\hat{\gamma}}}{\hat{\gamma}-1}\frac{D}{Dt}\left(\frac{P}{\rho^{\hat{\gamma}}}\right) = 0 \tag{4}$$

$$\frac{\partial\boldsymbol{B}}{\partial t} = \nabla\times(\boldsymbol{V}\times\boldsymbol{B}) \tag{5}$$

$$\nabla\cdot\boldsymbol{B} = 0 \tag{6}$$

$$p = \rho RT \tag{7}$$

where $\hat{\gamma}$ is the complex ratio of specific heats.

We take perturbation in the MHD equations of the form $exp\ i(\omega t+\boldsymbol{k}\cdot\boldsymbol{r})$ where ω is the frequency of oscillations and $\boldsymbol{k}$ is the wave number**.** The effect of Newtonian radiation can be incorporated by using an adiabatic energy equation with complex $\hat{\gamma}$ (Bunte and Bogdan, 1994), which is given by

$$\hat{\gamma} = \frac{1+\mathrm{i}\omega\tau_{\mathrm{R}}\gamma}{1+\mathrm{i}\omega\tau_{\mathrm{R}}}\ . \tag{8}$$

We can generalize the ideal MHD case by replacing sound speed the $C_s$ by a complex sound speed $\hat{C}_s$, given by

$$\hat{C}_s^2 = \frac{\hat{\gamma} p_0}{\rho_0} \tag{9}$$

where $\hat{\gamma} = \gamma$ in the adiabatic limit and $\hat{\gamma} = 1$ in the isothermal limit.

We cast the equations in dimensionless form as follows:

$$x = l\overline{x}, \quad \overline{k} = lk, \quad V = \mathrm{C_s}\overline{V}, \quad \omega = \frac{C_s}{l}\overline{\omega}, \quad B = B_0\overline{B}_x, \quad p = \frac{B_0^{\,2}}{\mu}\overline{p}, \quad \rho = \frac{\gamma B_0^{\,2}\overline{\rho}}{\mu C_S^{\,2}}$$

where $l$ is a dimensionless length and the quantities with a bar on top denote non-dimensional quantities.

Linearising the MHD equations (2-7), we obtain a fourth order dispersion relation as

$$a_4 k^4 + a_3 k^3 + a_2 k^2 + a_1 k + a_0 = 0 \tag{10}$$

where the coefficients in the dimensionless form are given by

$$a_0 = \omega^5 - \frac{\mathrm{i}\omega^4}{\tau_R}$$

$$a_1 = 0$$

$$a_2 = -\omega^3(1 + V_A^2) + \frac{i\omega^2}{\tau_R}\left(\frac{1}{\gamma} + V_A^2\right)$$

$$a_3 = 0$$

$$a_4 = \cos^2\theta\, V_A^2 \omega - \frac{i\cos^2\theta\, V_A^2}{\gamma\tau_R}$$

To examine spatial damping of waves, we take $\omega$ to be real and $k = k_R + ik_I$ as the complex wave number. The dispersion relation has four roots, corresponding to damped magnetoacoustic modes. The roots of the dispersion relation have been obtained numerically by using values of the physical parameters applicable to prominences as discussed in Terradas *et al.* (2001). We calculate the damping length, $d_L = \left(\frac{1}{k_I}\right)$ for different cases. The imaginary part of the complex wave number gives the damping length, $d_L$ for the slow and fast MHD modes. The slow mode is absent when $k_x$=0. The observed upper limit for the perturbation wave length is obtained by Molowny-Horas *et al.* (1997), which is 20 Mm. The dispersion relation gives a higher damping length for fast mode wave compared to slow mode. The manners in which radiative losses from the prominence medium will take place will be the same for slow and fast mode as is evident from figure 1 and 2.

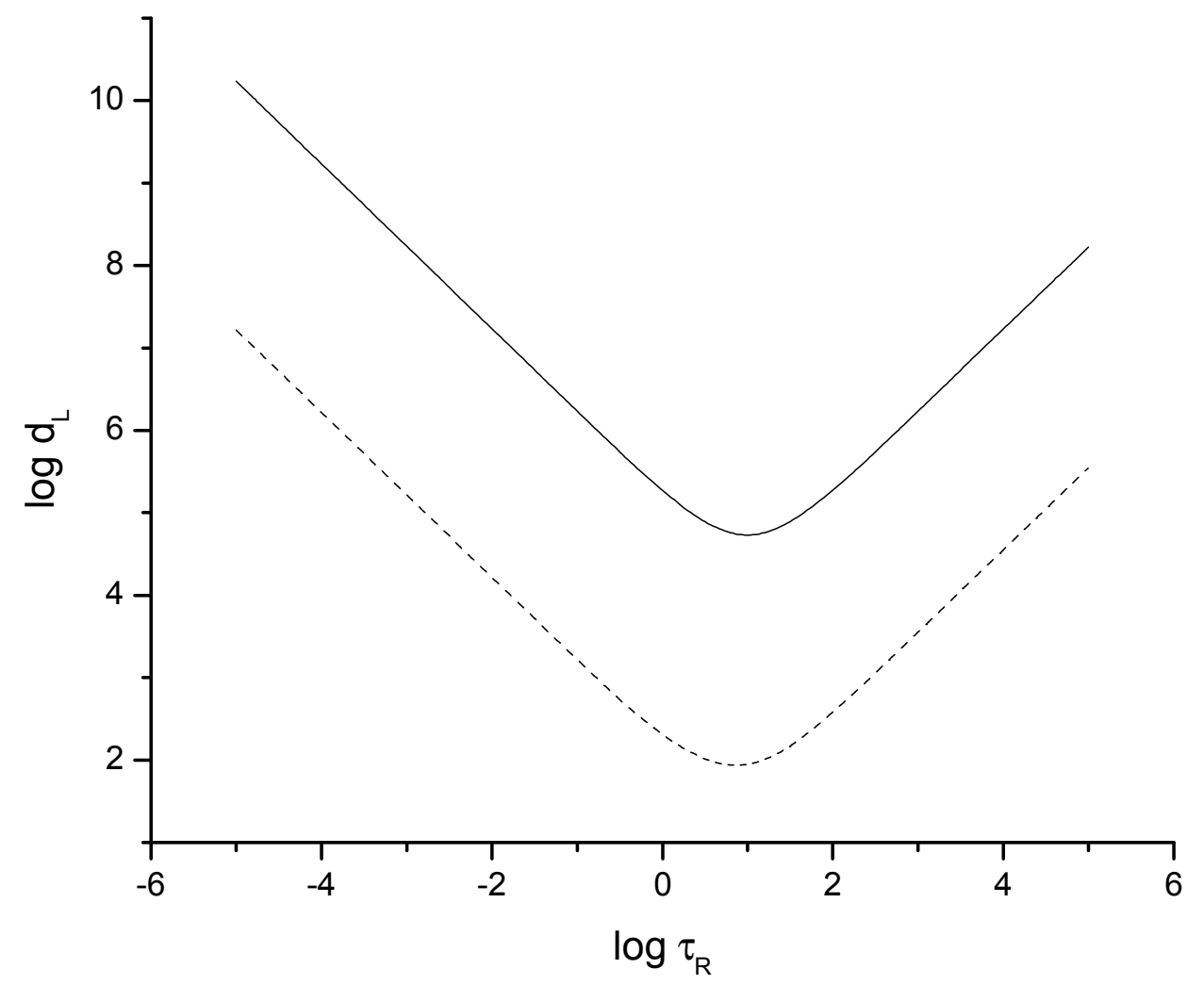

Figure. 1a. Damping length, $d_L$ as a function of the radiative relaxation time $\tau_R$ for MHD waves with $\omega = 10^{-1}$. (dotted line: slow mode, solid line: fast mode)

For a given value of frequency $\omega$, the damping length first decreases as a function of $\tau_R$ and attains a minimum then increases as we vary $\tau_R$ from $10^{-5}$ to $10^{5}$ (in dimensionless units) (see Figure. 1a and 1b). For $\tau_R \to \infty$, the wave takes infinite time to damp and therefore travels very long distances. This is due to the fact that radiative losses become almost inefficient for large radiative relaxation times. For $\omega = 10^{-1}$ and $\tau_R$ in the range $10^{-5}$ to $10^{5}$, significant damping of linear compressional waves is obtained for $\tau_R$ = 10, which corresponds to $\tau_R = 4.34\times10^{3}$ s as shown in Figure 1a.

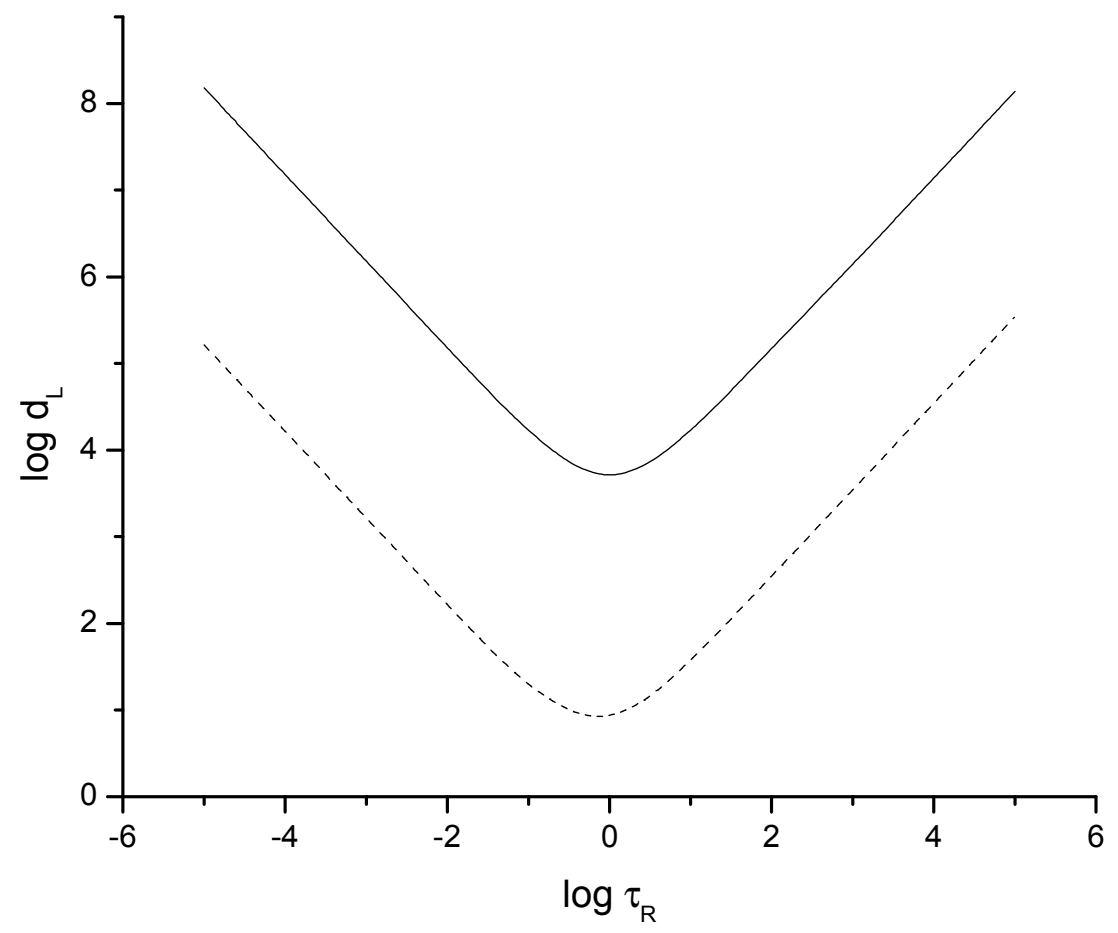

Figure. 1b. Damping length, $d_L$ as a function of the radiative relaxation time $\tau_R$ for MHD waves with ω = 1. (dotted line: slow mode, solid line: fast mode)

For a given value of the radiative relaxation time, the damping length decreases almost linearly with increasing ω (Figure 2a and 2b).

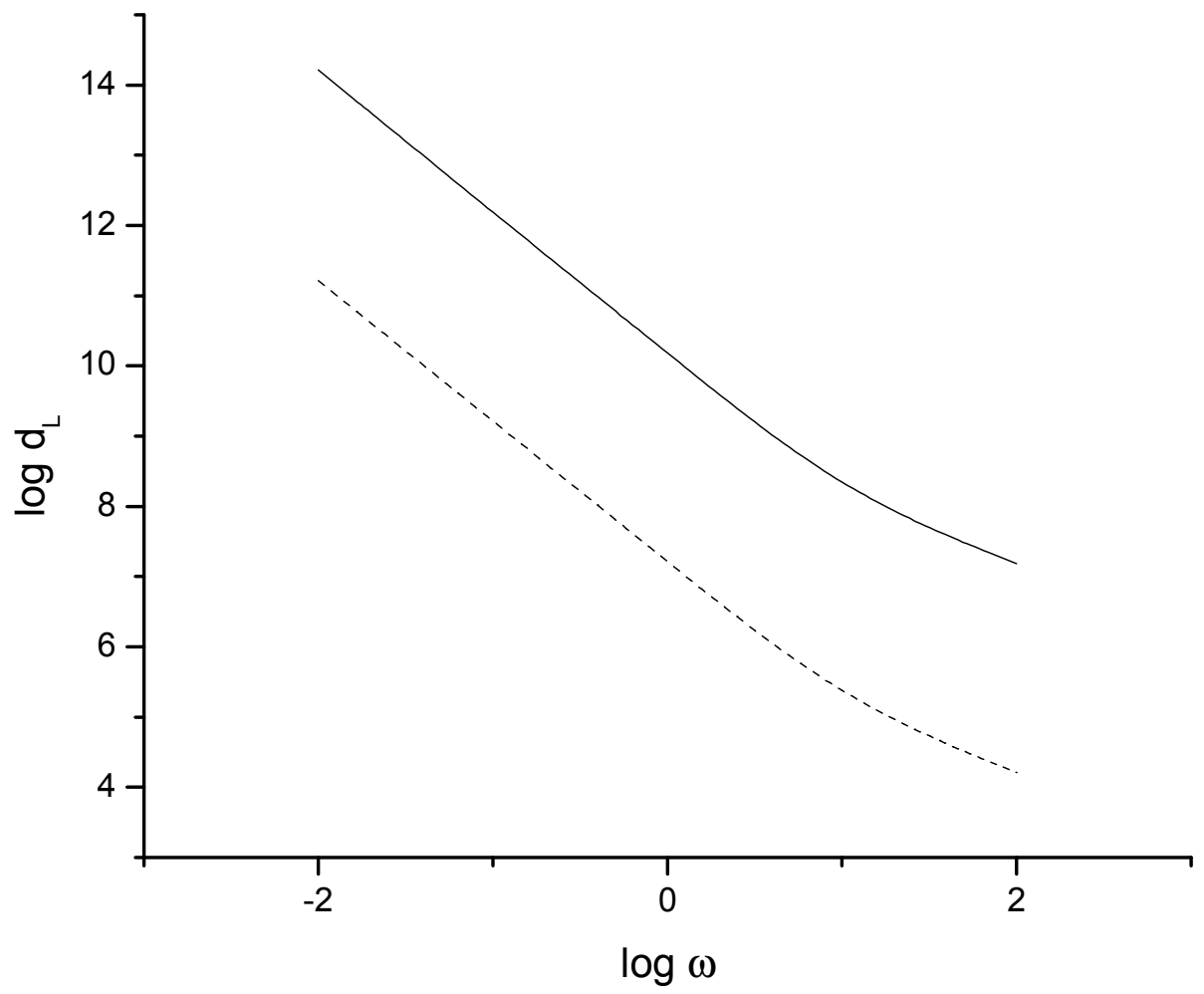

Figure. 2a. Damping length, $d_L$ as a function of the ω for MHD waves with $\tau_R = 10^{-7}$. (dotted line: slow mode, solid line: fast mode)

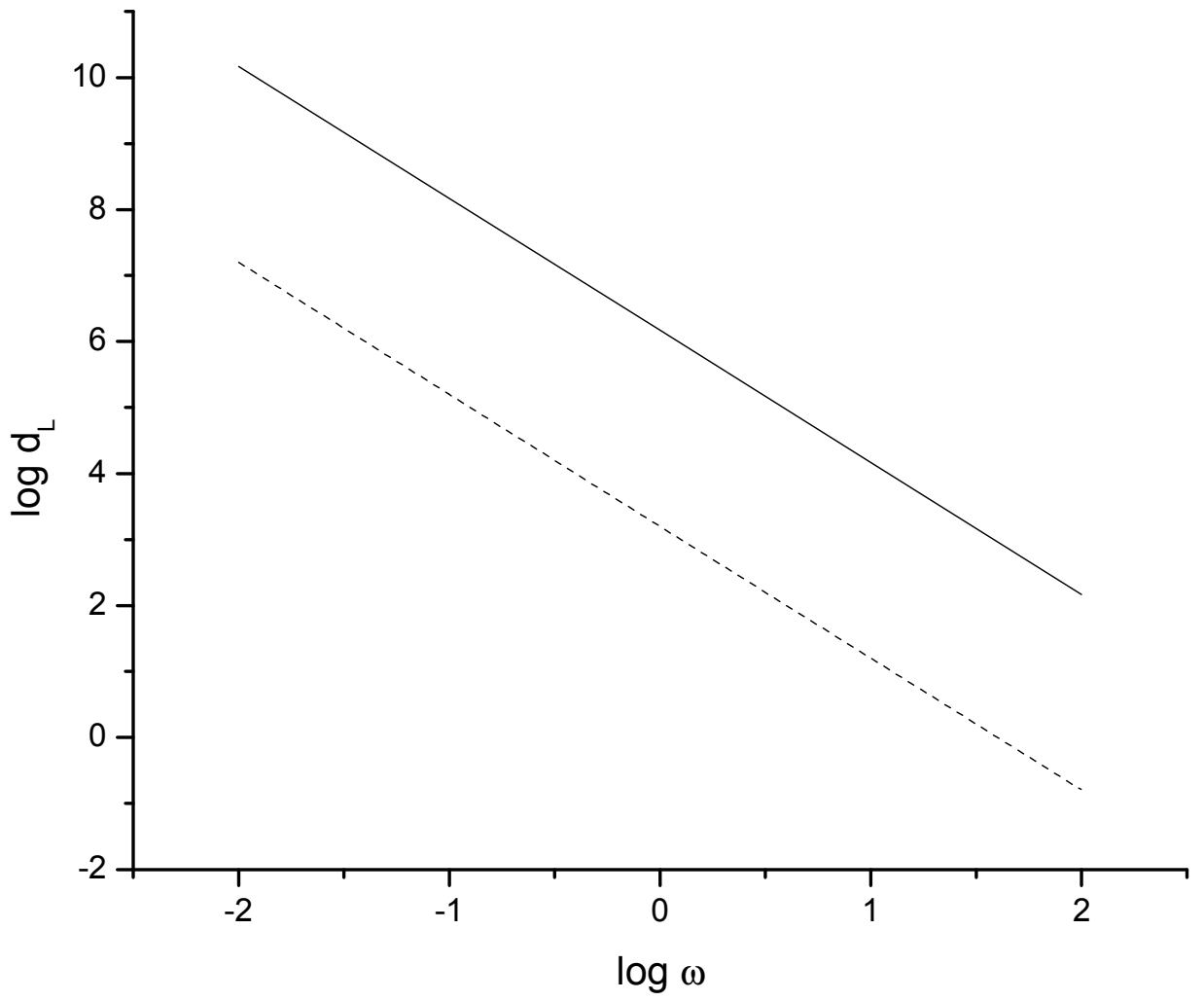

Figure 2b. Damping length, $d_L$ as a function of the ω for MHD waves with $\tau_R = 10^{-3}$. (dotted line: slow mode, solid line: fast mode)

Increasing the value of radiative relaxation time leads to a decrease of the damping length and at high frequencies both wave modes tend to have short damping lengths (Figure 2a and 2b). Therefore, high frequency MHD waves are highly damped.

We here discuss the method of prominence seismology for the determination of the opacity in the prominence that is based on the observation of damped oscillations in prominences by Molowny-Horas *et al.* (1999). Using the VTT telescope at Sac Peak, Molowny-Horas *et al.* (1999) found velocity perturbations with periods between 28 and 95 min at different locations in a prominence and observed that the amplitude of the oscillations decreases in time with damping times between 101 and 377 min. Terradas *et al.* (2001) considered the energy losses through Newtonian cooling and found that the slow mode waves are affected by damping leaving fast mode waves almost undamped. Interpretation of the observation of damped oscillations by the energy losses through Newtonian cooling allows us to connect the damping time of the waves and the opacity in prominences. The radiative relaxation time $\tau_R$ and solar opacity are related as (Spiegel, 1957)

$$\tau_R = \frac{C_v}{16\sigma_B\, \kappa T^3}, \qquad (11)$$

where $C_v$ is the volumetric specific heat, $\kappa$ is the opacity and $\sigma_B$ is the Stefan-Boltzmann constant. Taking $\tau_R = 6 \times 10^3$ s, T= $7 \times 10^3$ K (typical for prominences) the opacity $\kappa = 7.4 \times 10^{-4}$ m$^2$ kg$^{-1}$ in a prominence.

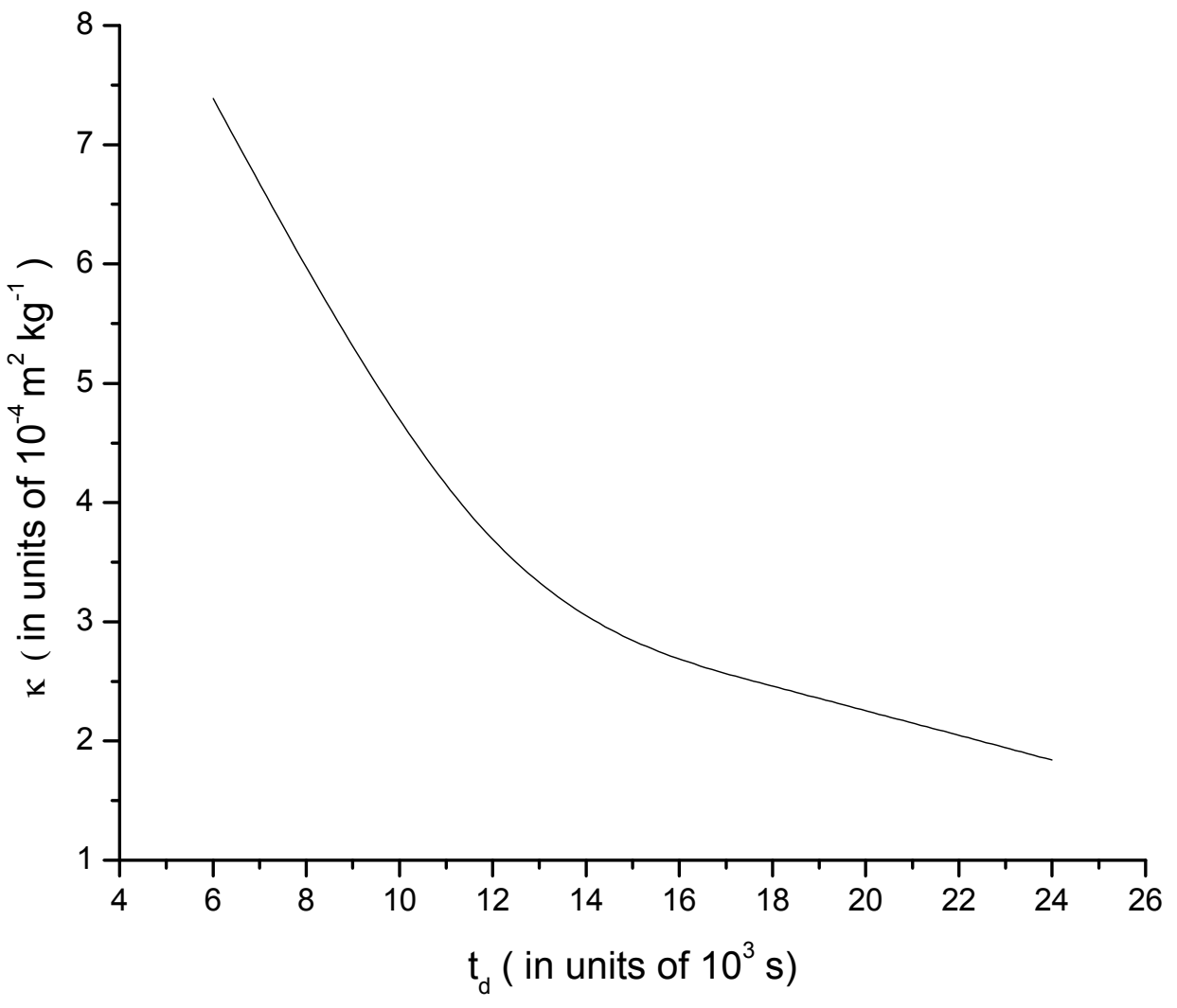

Figure 3a. Opacity as a function of damping time.

The upper observed limit for the perturbation wave length is obtained by Molowny-Horas *et al.* (1997). We have examined the effects of radiative losses due to Newtonian cooling for the spatial damping of linear compressional waves in quiescent prominences and prominence-corona transition region (PCTR). We have compared the results with the observed upper limit for the perturbation wave length obtained by Molowny-Horas *et al.* (1997), which is 20 Mm. The dispersion relation gives a higher damping length for fast mode wave compared to slow mode. This could be due to the fact that fast mode wave velocity $V_A$, is in the Alfvénic range for $\beta < 1$ plasma. The Alfvén wave, being incompressional in nature, is extremely difficult to dissipate in linear regime. Therefore, it is likely that the fast mode wave can be damped over a much larger distance as compared to the slow mode wave. It has been found that for a given value of radiative relaxation time, the high frequency slow mode waves are highly damped. Interpretation of upper observed limit to perturbation wavelength as the damping length and observed damping of waves due to the energy losses through Newtonian cooling allows us to connect the opacity with the damping length of the waves in the prominences. The opacity is related to the damping length $l_d$ as follows:

$$\kappa = \frac{C_s C_v}{16 \sigma_B l_d T^3} . \qquad (12)$$

Considering the upper observed limit to the perturbation wavelength in prominences (Molowny-Horas *et al.*, 1997) as $l_d$= 20 Mm we have $\kappa$=2.6 $\times 10^{-3}$ m$^2$ kg$^{-1}$. The opacity is inversely proportional to the damping time and damping length.

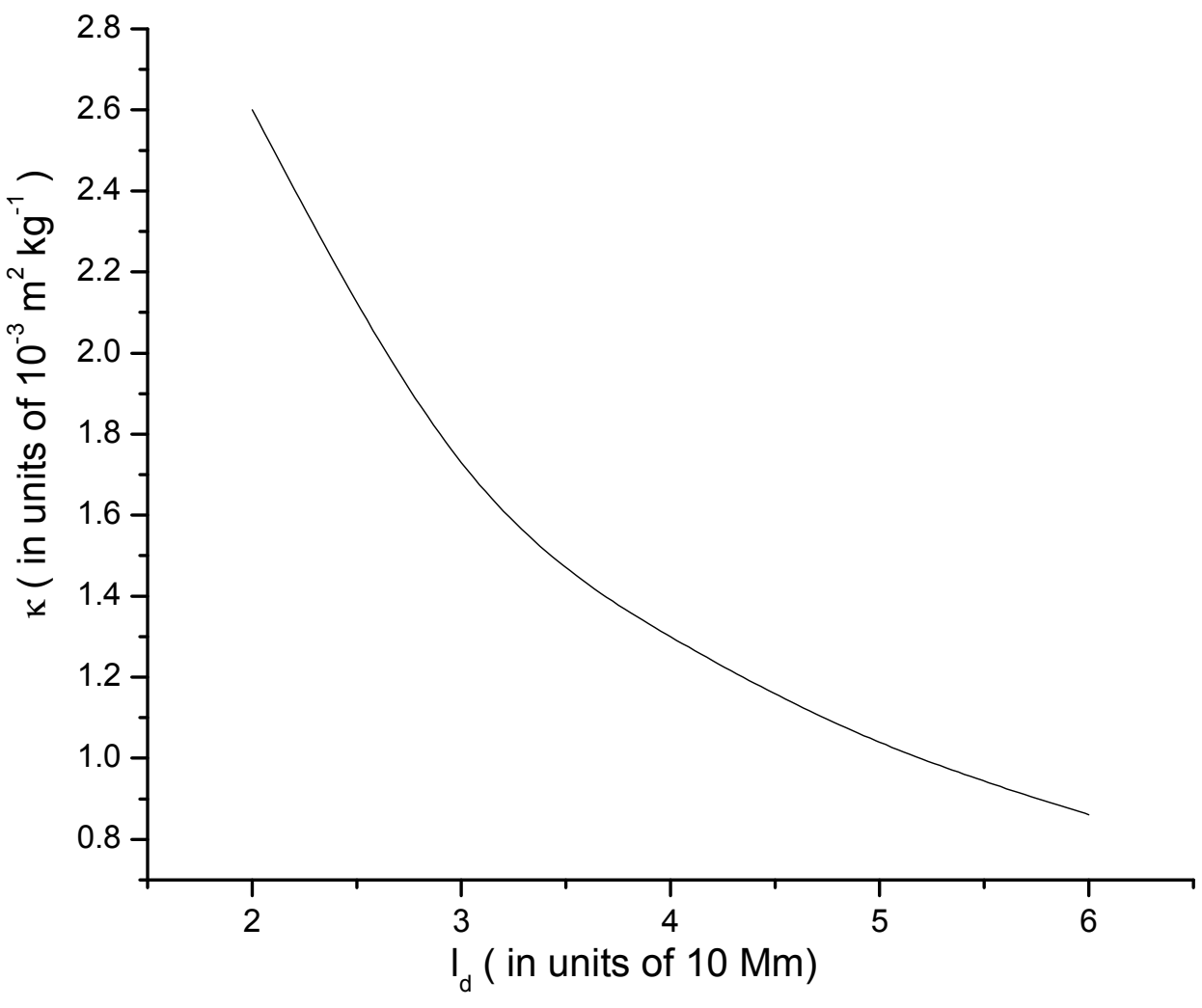

Figure 3b. Opacity as a function of damping length.

As the damping time increases the opacity in a prominence linearly decreases (Figure 4a). Increasing the value of damping length decreases the opacity (Figure 4b). Using prominence seismology, the opacity has been calculated taking account of the damping time and the upper observed limit to the perturbation wavelength. It turns out that the opacity values calculated using damping times are orders of magnitude different compared to that using the damping length. The value of opacity calculated using slightly higher value of the damping length (60 Mm) can give the same order of magnitude results compared to that calculated using damping time.

The value of the radiative relaxation time $\tau_R$ in prominences is not known. Let us introduce a radiative energy loss term $L_r$ in eq. (4) of the form

$$L_r = \rho^2 \chi^* T^\alpha ,$$

where $\chi^*$ and $\alpha$ are constants and the radiative relaxation time $\tau_R$ is given by

$$\tau_R = \frac{\gamma p}{(\gamma - 1)\chi^* \rho^2 T^\alpha} ,$$

which for different prominence regimes mentioned in Table 1 (Carbonell *et al.*, 2004) is between $10^2$-$10^4$s both for prominence and the PCTR. The theoretical value of radiative relaxation time in the corona typically is $6\times10^4$ s (Priest *et al.*, 1991). The value of radiative relaxation time calculated using opacity values is $1.1 \times 10^{-4}$s (Ballai, 2003). Our calculations, using the observed value to fix an upper limit to perturbation wavelength clearly indicates that the value of radiative relaxation time in prominences lies between 10-$10^3$ s. Also, the inclusion of more realistic radiative term in the energy equation gives almost similar range of damping lengths.

## 3. Role of Magnetohydrodynamic Turbulence

The presence of MHD turbulence in quiescent prominences may give rise to chaotic velocity fields (Jensen, 1982, 1983). The damping of MHD waves due to turbulent phenomenon has been investigated in coronal loops (Nakariakov *et al.*, 1999). In some cases, large turbulent velocities have also been observed in the corona close to quiescent prominences (Jensen, 1982). The micro-turbulent velocity in the range 2-8 $\mathrm{kms}^{-1}$ has also been observed in prominences (De Boer *et al.*, 1998). Non-thermal velocities up to 36 $\mathrm{kms}^{-1}$ have been measured from SUMER/SOHO observations and length scales of 15 km and 1000 km have also been inferred respectively for turbulence characterized by Kraichnan and Kolmogorov distributions (Cirigliano *et al.*, 2004). The presence of turbulence can in principle enhance the transport coefficients like viscosity and thermal conductivity. We assume that the turbulence is due to Alfvén waves entering from below, and propagating along the magnetic field structures which are anchored in the photosphere (Jensen, 1983).

The observation of damped oscillatory motions in prominences by Molowny-Horas et al. (1997) and the hypothesis of MHD turbulence allow us to connect the turbulent viscosity and the damping time. We define the viscous damping time $t_d$ due to magnetohydrodynamic turbulence as

$$t_d = \frac{L^2}{\nu_t} \qquad (13)$$

where $L$ is the characteristic length scale of the prominence and $\nu_t$ is the turbulent kinematic viscosity.

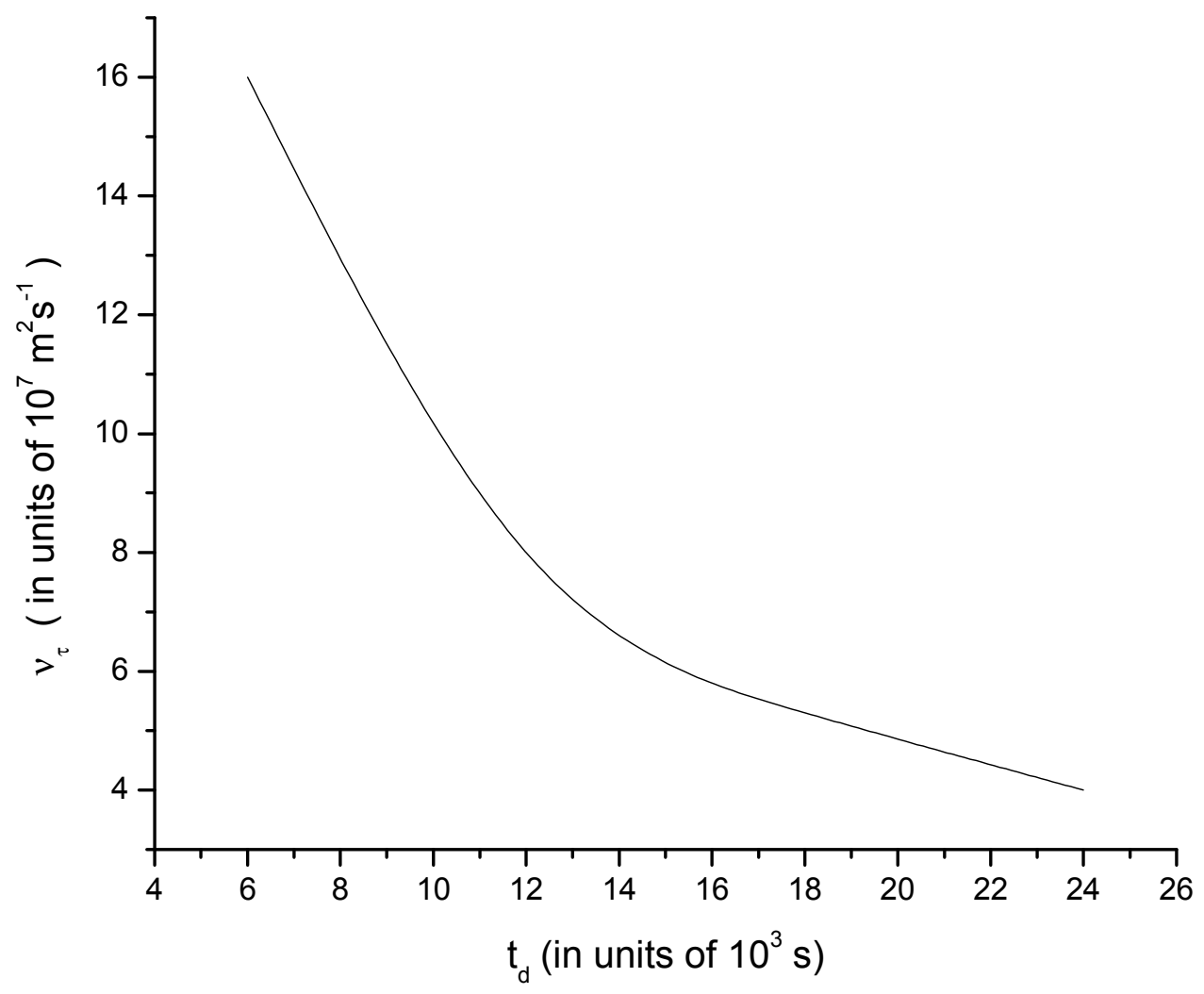

Figure 4a. Kinematic viscosity as a function of damping time.

For typical values of $t_d$ =6 $\times 10^3$ s and $L$ = 1 Mm, we have $\nu_t$ = 1.6 × $10^8$ $m^2$ $s^{-1}$. It is evident from the formula that the viscosity can be related to the size of the prominence and damping time. The classical value of kinematic viscosity in prominences typically is 2.8 $\times 10^5$ $m^2$ $s^{-1}$ (Ballai, 2003). Therefore, the calculated viscosity value using prominence seismology is orders of magnitude higher than the classical one. We attribute this to the presence of due to turbulence in prominences.

Interpretation of the upper observed limit to perturbation wavelength as the damping length and the hypothesis of MHD turbulence allow us to connect the turbulent viscosity with the damping length. The turbulent kinematic viscosity can be computed taking account of the upper observed limit to the perturbation wavelength (Molowny-Horas *et al.*, 1997) by using the formula

$$\nu_t = \frac{C_s L^2}{l_d} \qquad (14)$$

with $l_d$ = $C_s t_d$ where $l_d$ is the damping length for the slow mode wave and $C_s$ is the sound speed.

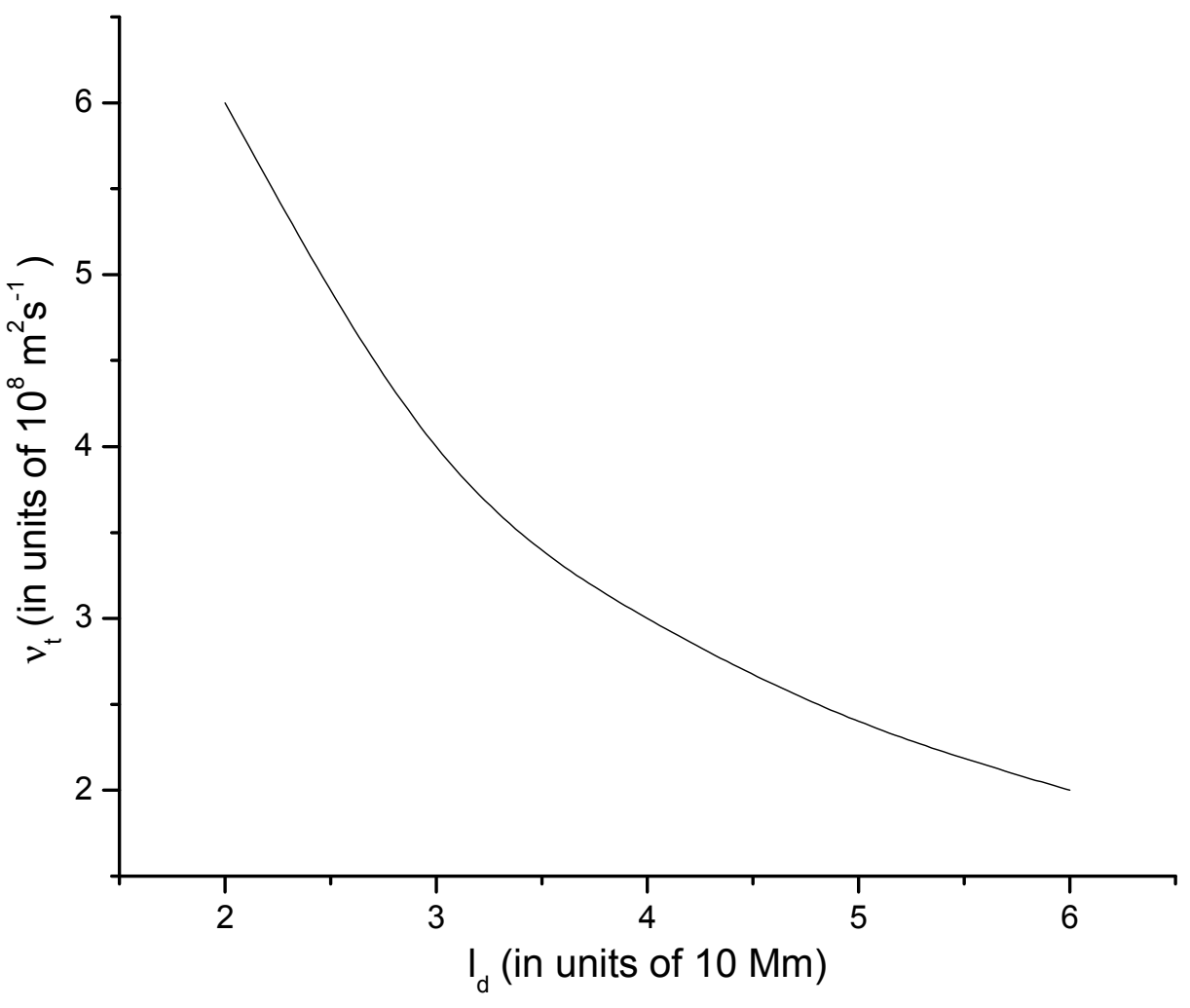

Figure 4b. Kinematic viscosity as a function of damping length.

The value of turbulent kinematic viscosity calculated using damping length differs by some orders of magnitude with the one calculated using damping time. This is because of the fact that turbulent kinematic viscosity depends upon damping length/damping time and on the characteristic length scale $L$ of the prominence.

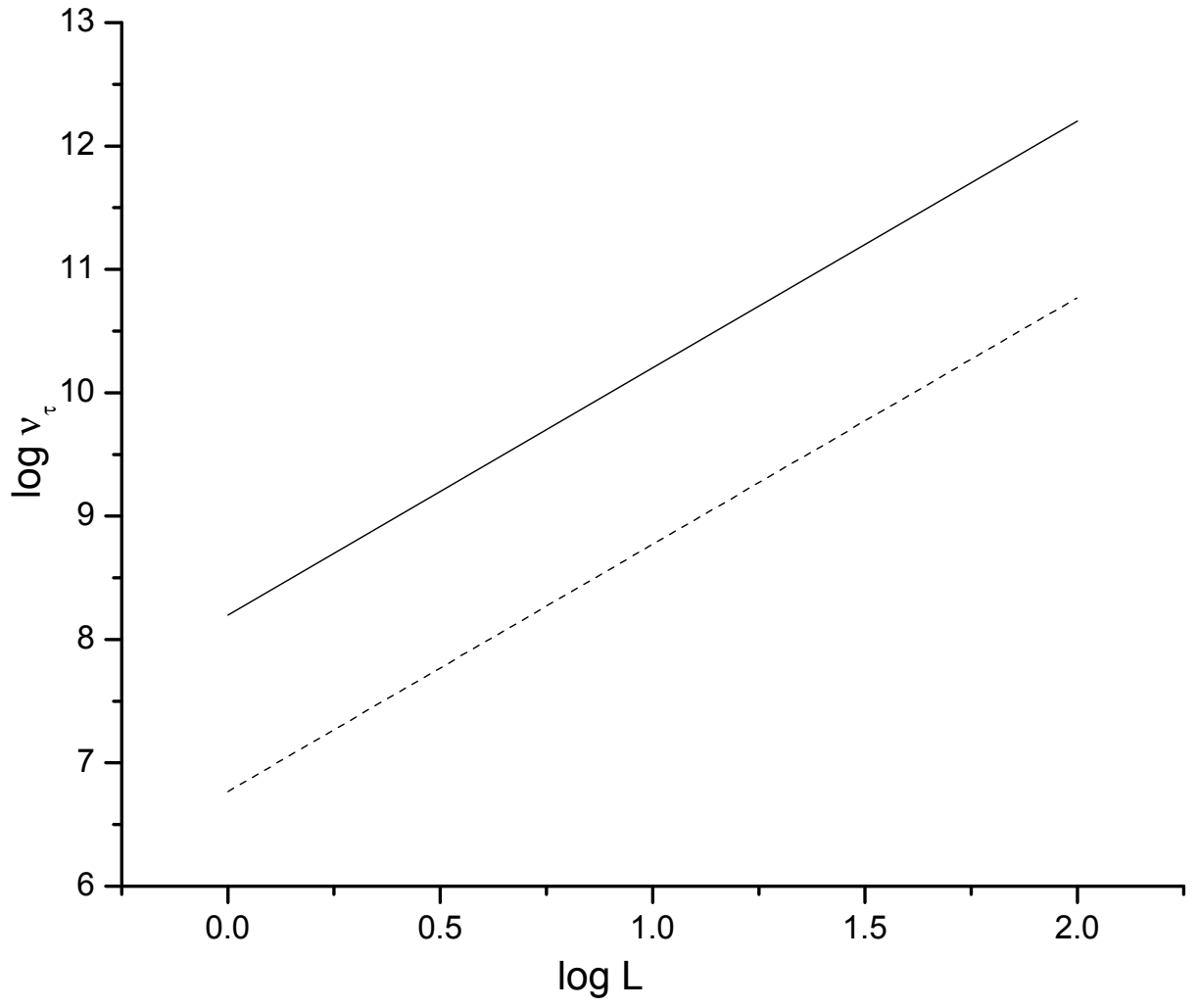

Figure 4c. Kinematic viscosity as a function of characteristic length scale of the prominence. (dotted line: using damping length, solid line: using damping time)

For typical values of the sound speed $C_s$ ~12 km s$^{-1}$ and $L = 10^3$ km for prominences we have $\nu_t = 6 \times 10^6$ m$^2$ s$^{-1}$. The value of turbulent kinematic viscosity will be even larger if we take $C_s$ ~90 km s$^{-1}$ (Cirigliano *et al.*, 2004). In this case the value of turbulent kinematic viscosity may be as large as $4.9 \times 10^9$ m$^2$ s$^{-1}$.

The interpretation of non-thermal motions in terms of MHD turbulence is attractive because all non-thermal motions are isotropic and have very small scales that are often displayed on the basis of their observational characteristics and the physical conditions in the prominence, PCTR, CCTR and the solar corona that may exhibit turbulence. The inhomogeneous structures caused by density stratification and the magnetic field may also play a role in driving turbulence (Chae *et al.*, 1998). The presence of non-thermal velocities and damping of magnetoacoustic waves could, therefore, be explained by using the hypothesis of MHD turbulence. We have evaluated the required value of the turbulent kinematic viscosity in prominences which can explain the damping in prominences. Cirigliano *et al.* (2004), from studies of non-thermal velocity profile found similarity with CCTR which indicated that the heating mechanisms in PCTR could be same as in the CCTR as far as wave propagation and MHD turbulence is considered. Chae *et al*. (1998), from non-thermal velocity measurements, deduced the importance of Alfvén wave heating in CCTR and also considered MHD turbulence mechanisms to explain non-thermal velocities. De Boer *et al*. (1998) have calculated non-thermal velocities and have found these to increase towards the periphery of the prominence. Terradas *et al*. (2002), from temporal and spatial analysis observed this common feature in large areas, especially close to the edge of the prominence. The orders of magnitude greater viscosity value, in our calculations, in PCTR is in agreement to their results.

## 4. Results and Discussion

The observations of the time series of Doppler widths, line width and line intensity clearly shows that the prominence oscillates with the period between few minutes to few hours. One of the typical features of these oscillations is the damped oscillatory motions in prominences. Molowny-Horas et al. (1997) recently observed oscillations in different parts of a quiescent prominence and assuming plane wave propagation, they obtained the group speed, phase speed and the perturbation wavelength. Molowny-Horas *et al*. (1999), using the VTT telescope of Sac Peak Observatory, found velocity perturbations with periods between 28 and 95 min at different locations in a prominence and observed that the amplitude of the oscillations decreases in time with damping times between 101 and 377 min. Terradas et al. (2002) investigated the temporal and spatial variations of the oscillations and reported the strong damping of the oscillations with the damping times between two to three times the wave period. The physical mechanism that works behind the damping of these oscillations is not known.

Terradas et al. (2001), on the theoretical side, modeled the damped oscillatory motions as the energy losses due to the Newtonian cooling and found that the slow mode waves are heavily damped, whereas fast modes are practically unaffected by the radiative dissipation and have very long damping times. We have examined the effects of radiative losses due to Newtonian cooling and MHD turbulence for the spatial damping of linear compressional waves in quiescent prominences and PCTR and studied using the local dispersion relation method. The dispersion relation gives a higher damping length far fast mode wave compared to slow mode. This could be due to the fact that fast mode wave velocity $V_A$, is in the Alfvénic range for $\beta < 1$ plasma. The Alfvén wave, being incompressional in nature, is extremely difficult to dissipate in linear regime. Therefore, it is likely that the fast mode wave can be damped over a much larger distance as compared to the slow mode wave. For a given value of frequency $\omega$, the damping length first decreases as a function of $\tau_R$ and attains a minimum then increases as we vary $\tau_R$ from $10^{-5}$ to $10^{5}$ (in dimensionless units) (see Figure. 1a and 1b). For $\tau_R \rightarrow \infty$, the wave takes infinite time to damp and therefore travels very long distances. This is due to the fact that radiative losses become almost inefficient for large radiative relaxation times. For $\omega = 10^{-1}$ and $\tau_R$ in the range $10^{-5}$ to $10^{5}$, significant damping of linear compressional waves is obtained for $\tau_R = 10$, which corresponds to $\tau_R = 4.34\times10^{3}$ s as shown in Figure 1a. For a given value of the radiative relaxation time, the damping length decreases almost linearly with increasing $\omega$ (Figure 2a and 2b). Increasing the value of radiative relaxation time leads to a decrease of the damping length and at high frequencies both wave modes tend to have short damping lengths (Figure 2a and 2b). Therefore, high frequency MHD waves are highly damped. The damping length has been plotted as a function of the frequency and radiative relaxation time (Figure 5).

There are various mechanisms such as Spitzer's viscosity, thermal and electrical conduction and radiative cooling that can, in principle, be responsible for the damping of the MHD waves. Recently, Ballai (2003) has reviewed some of the possible mechanisms that operate in the coronal prominences that can explain the spatial damping of the prominences. The viscosity in prominence plasma is mainly due to the ions and the electron contribution to it is negligible. Ballai (2003), in his order of magnitude calculation, showed that the Spitzer's (or classical) viscosity does not play an important role in the damping of the magnetoacoustic waves. In addition to this, the viscosity in prominence plasma was found to be mainly isotropic. However, the thermal conduction which is mainly due to the electrons is anisotropic in nature. Ballai (2003) concluded that the thermal conduction can be a viable damping mechanism provided the wavelength of the waves is short and for a wave period of 40 min the thermal conduction can be safely neglected. Carbonell *et al.* (2004) studied the damping of MHD waves in an unbounded medium by considering the effect of prominence-corona transition region (PCTR) with the energy losses due to thermal conduction, optically thin radiation and heating. They found out that the damping times of fast modes are very long compared to the slow modes. Our result clearly shows that the damping lengths of fast modes are very long compared to the slow modes. Terradas *et al*. (2005) showed that the result also holds for the bounded medium. The different heating mechanisms that were applied to different prominence regimes do not produce appreciable change on the damping profiles (Carbonell *et al*., 2004).

The radiative relaxation time in prominences is not known due to the lack of the opacity measurements. Using the fact that the damping of slow mode waves is caused by the energy losses due to the Newtonian cooling we have evaluated the opacity in the prominence plasma. Equation (11) and (12) show that the opacity can be evaluated in the prominence plasma provided the physical mechanism behind the damping is known. We have evaluated the opacity using the observed values of the damping time (Figure 3a) and the damping length (Figure 3b). It turns out that the opacity values calculated using these two formulas have orders of magnitude difference. The value of opacity calculated using slightly higher value of the damping length (60 Mm) can give the same order of magnitude results compared to that calculated using the damping time.

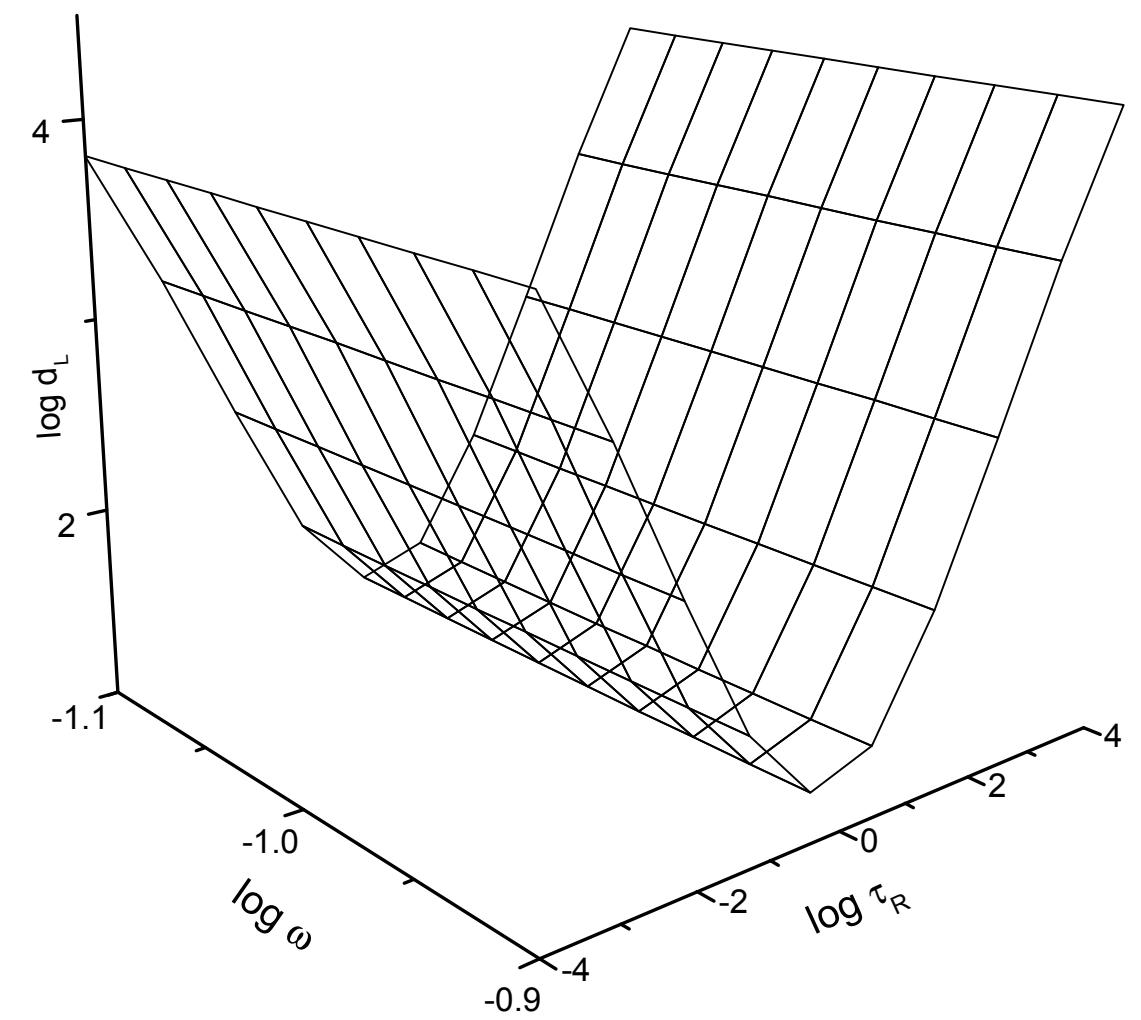

Figure 5a. Slow mode damping length as a function of frequency and radiative relaxation time.

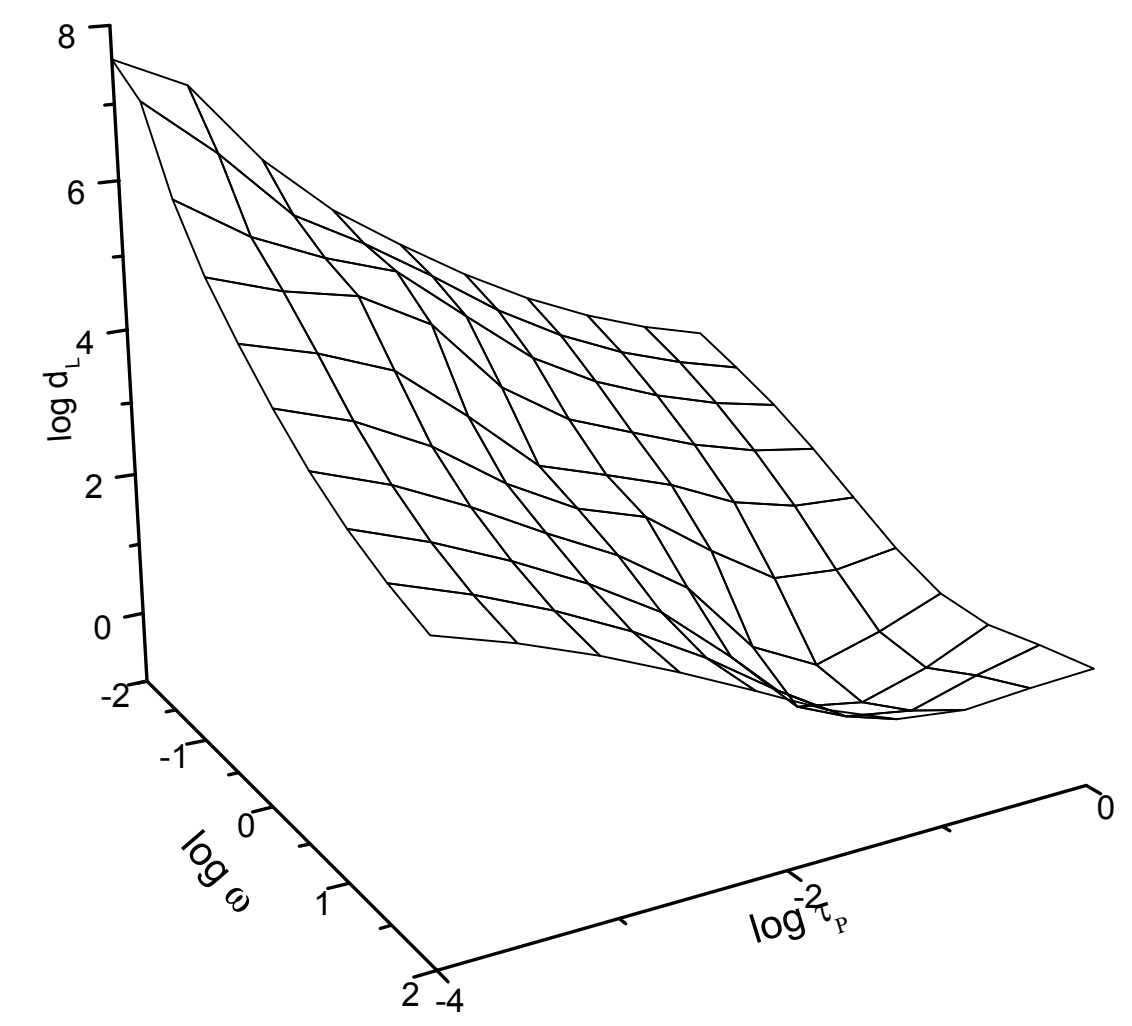

Figure 5b. Slow mode damping length as a function of frequency and radiative relaxation time.

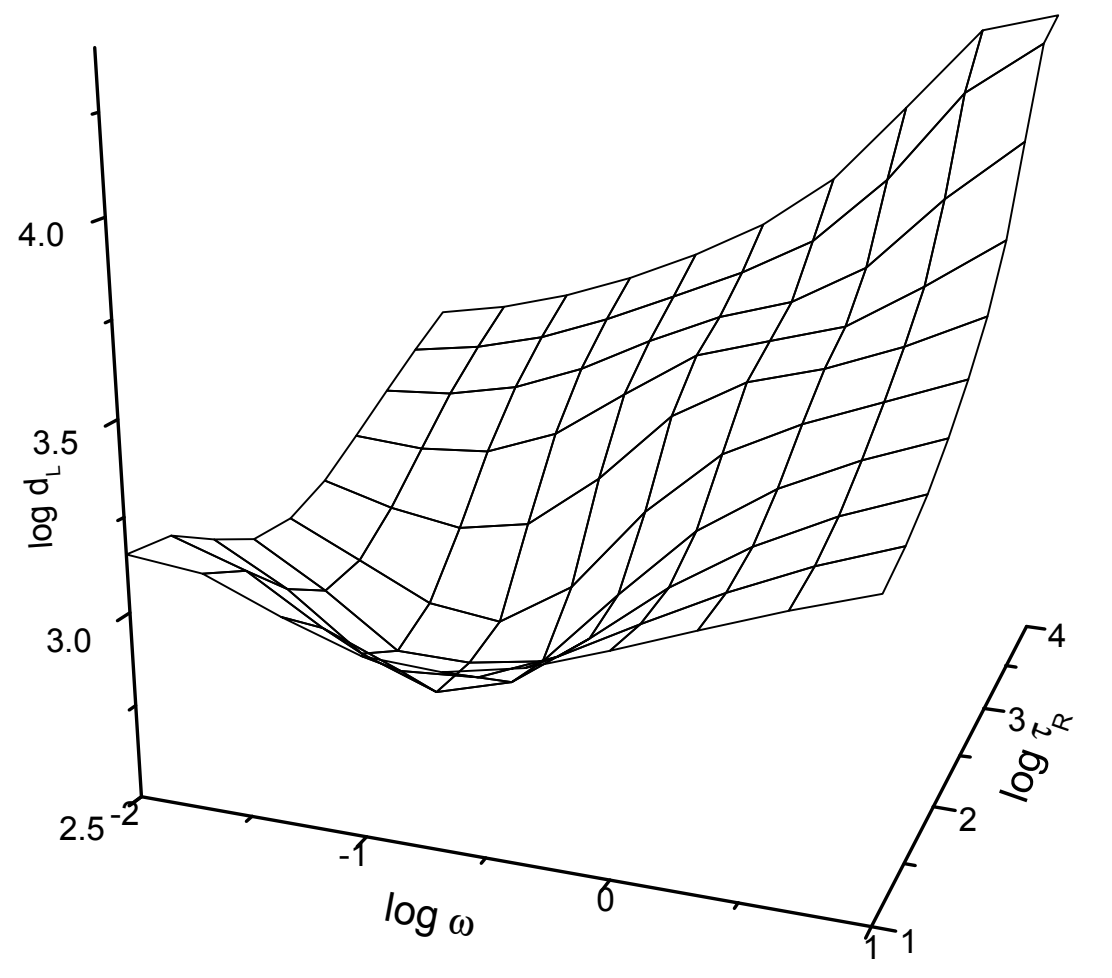

Figure 5c. Slow mode damping length as a function of frequency and radiative relaxation time.

The energy losses through Newtonian cooling alone are inadequate to explain the damping of fast mode waves but gives acceptable damping lengths for the slow mode using radiative relaxation times in the range $10\text{-}10^3$ s. Therefore, it is important to look for an alternative physical mechanism that can explain the damping of both slow and fast mode wave. The role of MHD turbulence in damping of linear compressional MHD waves cannot be ruled out in the light of SUMER observations of non-thermal velocity and length scale. We have calculated the value of turbulent kinematic viscosity in prominence plasma that can reproduce the observed damping time and damping length in prominences (Figure 4a and 4b). The dependence of the turbulent kinematic viscosity on the characteristic length scale is clearly shown in Figure 3c. The turbulent kinematic viscosity has been calculated using the SUMER observations of the sound speed in the PCTR. The turbulent kinematic viscosity in the PCTR is larger than the main body of the prominence. Also it is found that the turbulent kinematic viscosity is orders of magnitude higher than the classical one. This is due to the fact that the turbulence enhances the transport coefficients like viscosity and thermal conductivity. The convective disturbances that can propagate along the magnetic field structures which are anchored in the photosphere may cause the MHD turbulence in the prominences. The inhomogeneous structures caused by the density stratification and the magnetic field may increase complexity and play a role in driving turbulence (Chae *et al*., 1998).

In conclusion, we find that the radiative losses give acceptable damping lengths for the slow mode wave for the radiative relaxation times in the range ($10\text{-}10^3$s). From prominence seismology, the values of opacity (Figure 3) and turbulent kinematic viscosity (Figure 4) have been inferred. It has been found that for a given value of the radiative relaxation time, the high frequency slow mode waves are highly damped. We have investigated the possible role of MHD turbulence in damping of MHD waves and found a turbulent viscosity can re-produce the observed damping time and damping length in prominences, especially in PCTR. The presence of density inhomogenieties may lead to phase mixing and resonant absorption of MHD waves, which also needs to be investigated.

## Acknowledgements

KAPS acknowledges the UGC, New Delhi for the award of a research fellowship and the hospitality of the Indian Institute of Astrophysics, Bangalore where a part of the works was carried out. This work was also benefited from detailed comments from S.S. Hasan, B.N. Dwivedi, India and R. Oliver, Spain. We are thankful to the anonymous referee for his valuable suggestions for the improvement of the paper.

**References**

Ballai, I.: 2003, Astron. Astrophys. **410**, L17.

Bashkirtsev, V.S. and Mashnich, G.P.: 1984, Solar Phys. **91**, 93.

De Boer, C.R., Stellmacher, G., and Wiehr, E.: 1998, Astron. Astrophys. **334**, 280.

Bunte, M., and Bogdan, T.J.: 1994, Astron. Astrophys. **283**, 642.

Carbonell, M., Oliver, R., and Ballester, J.L.: 2004, Astron. Astrophys. **415**, 739.

Chae, J., Schühle, U., and Lemaire, P.: 1998, Astrophys. J. **505**, 957.

Cirigliano, D., Vial. J.C., and Rovira, H.: 2004, Solar Phys. **223**, 95.

Jensen, E.: 1982, Solar Phys. **77**, 181.

Jensen, E.: 1983, Solar Phys. **89**, 275.

Molowny-Horas, R., Oliver, R., Ballester, J.L. and Baudin, F.: 1997, Solar Phys. **172**, 181.

Molowny-Horas, R., Wiehr, E., Balthasar, H., Oliver, R. and Ballester, J.L.: 1999, in *JOSO Annual Report '98*, ed. A. Antalova', H. Balthasar, and A. Kučera, Astronomical Institute Tatranská Lomnica, p. 126.

Nakariakov, V.M., Ofman, L., Deluca, E., Roberts, B., and Davila, J.M.: 1999, Science, **285**, 862.

Patsourakos, S. and Vial, J.C.: 2002, Solar Phys. **208**, 253.

Priest, E.R., Hood, A.W., and Anzer, U.: 1991, Solar Phys. **132**, 199.

Spiegel, E.A.: 1957, Astrophys. J. 1957, **126**, 2002.

Terradas, J., Carbonell, M., Oliver, R. and Ballester, J.: 2005, Astron. Astrophys. **434**, 741.

Terradas, J., Oliver, R., and Ballester, J.L.: 2001, Astron. Astrophys. **378**, 635.